\documentclass[aps,prb,twocolumn,superscriptaddress,showpacs]{revtex4-1}

\usepackage{graphicx}

\usepackage[english]{babel}
\usepackage{color}
\usepackage{amsmath}
\usepackage{mathtools}

\begin{document}



\title{Non-Gaussian tail in the force distribution:
 A hallmark of correlated disorder in the host media of elastic objects}

\author{Jazm\'{i}n Arag\'{o}n S\'{a}nchez}%
\author{Gonzalo Rumi}
\author{Ra\'ul Cort\'es Maldonado}%
\author{N\'estor Ren\'e Cejas Bolecek}%
\author{Joaqu\'{i}n Puig}%
\author{Pablo Pedrazzini}%
\author{Gladys Nieva}%
\affiliation{Centro At\'{o}mico Bariloche and Instituto Balseiro,
CNEA, CONICET and Universidad Nacional de Cuyo, 8400 San Carlos de
Bariloche, Argentina}
\author{Moira I. Dolz}
\affiliation{Universidad Nacional de San Luis and Instituto de
F\'{i}sica Aplicada, CONICET, 5700 San Luis, Argentina.}
\author{Marcin Konczykowski}%
\affiliation{Laboratoire des Solides Irradi\'{e}s, CEA/DRF/IRAMIS,
Ecole Polytechnique, CNRS, Institut Polytechnique de Paris, 91128
Palaiseau, France}
\author{Cornelis J. van der Beek}%
\affiliation{Centre de Nanosciences et de Nanotechnologies, CNRS,
Université Paris-Sud, Université Paris-Saclay, 91120 Palaiseau,
France.}
\author{Alejandro B. Kolton}%
\author{Yanina Fasano$^{*}$}%
\affiliation{Centro At\'{o}mico Bariloche and Instituto Balseiro,
CNEA, CONICET and Universidad Nacional de Cuyo, 8400 San Carlos de
Bariloche, Argentina}

\date{\today}

\begin{abstract}

Inferring the nature of disorder in the media where elastic objects
are nucleated is of crucial importance for many applications but
remains a challenging basic-science problem.  Here we propose a
method to discern whether weak-point or strong-correlated disorder dominates
based  on characterizing the distribution of the
interaction forces between objects mapped in large
fields-of-view. We illustrate our
 proposal with the case-study system of vortex structures nucleated
in type-II superconductors with different pinning landscapes.
Interaction force distributions are computed from individual vortex
positions imaged  in thousands-vortices fields-of-view in a
two-orders-of-magnitude-wide vortex-density range. Vortex structures
nucleated in point-disordered media present Gaussian distributions
of the interaction force components. In contrast, if the media have
dilute and randomly-distributed correlated disorder, these
distributions present non-Gaussian algebraically-decaying tails for
large force magnitudes.  We propose that detecting this deviation
from the Gaussian behavior is a fingerprint of strong disorder, in our
case originated from a dilute distribution of correlated pinning centers.

\end{abstract}


 \maketitle

 Elastic objects nucleated and driven in
disordered media represent an ubiquitous situation in nature,
covering diverse fields of research such as  defect and crack
nucleation and propagation in
materials,~\cite{Moretti2004,Ponson2017} domain wall
dynamics,~\cite{Brazovskii2004,Ferre2013,Paruch2013} charge density
waves,~\cite{Brazovskii2004,Gruner1988} photonic
solids,~\cite{Man2013}  randomly-packed objects,~\cite{Kurita2011}
magnetic bubbles nucleated in substrates,~\cite{Kulikova2016}
colloidal spheres,~\cite{Murray1990} to avalanches in
magnets~\cite{Urbach1995} and vortex matter in
superconductors.~\cite{Blatter1994,Giamarchi1995,Natterman2000} The
physical properties of these systems with different types of
particle-particle interaction has been the subject of a wide and
interdisciplinary field of
research.~\cite{Brazovskii2004,Ferre2013,Paruch2013,Gruner1988,Man2013,Kurita2011,Kulikova2016,Murray1990,Urbach1995,Blatter1994,Giamarchi1995,Natterman2000,LeDoussal2009,Fasano2005,Wu2009,Guyonnet2012,Dreyfus2015,Weijs2015}
Knowledge on the nature of disorder in the host media is of crucial
importance for many applications but in most cases requires
potentially destructive micro-structural characterization. Inferring
the nature of disorder via a non-invasive approach relying on
information from physical properties of the elastic objects remains
an open problem.

Research on  vortex matter nucleated in type-II superconductors  has
 shed light on the structural properties of elastic
objects nucleated in media with different types of
disorder.~\cite{Leghisa1993,Dai1994,Harada1996,Bezryadin1996,Troyanovski1999,
Fasano1999,Fasano1999b,Fasano2000,Grigorenko2001,Surdeanu2001,Silevitch2001,Field2002,Fasano2002,Menghini2003,vanBael2003,
Fasano2003,Veauvy2004,Karapetrov2005,Bjornsson2005,Yurchenko2006,Fischer2007,
Fasano2008,Petrovic2009,Suderow2014,AragonSanchez2019,Rumi2019,Llorens2020}
A handful of works study the spatial distribution of
particle-particle interaction forces to get information on the
vortex-disorder
interaction.~\cite{Demirdis2011,Yang2012,vanderBeek2012,Demirdis2013,Yaguil2016,
CejasBolecek2016,AragonSanchez2019b}  Superconducting vortices are
repulsively-interacting elastic lines but, due to the pressure
exerted by the applied field, they tend to form a hexagonal lattice
with a spacing $a_{0}$  tuned by the magnetic induction $B$, namely
$a_{0} \propto B^{-1/2}$. In addition, vortices are pinned by
disorder, structural defects naturally present or introduced in the
host superconducting samples. As in many systems of interacting
elastic objects, the structural properties of vortex matter result
from the balance between thermal, particle-particle and
particle-disorder interaction energies.  Many theoretical and
experimental studies describe the structural deviations from a
perfect hexagonal vortex lattice induced by different pinning
landscapes.~\cite{Blatter1994,Leghisa1993,Dai1994,Harada1996,Bezryadin1996,Troyanovski1999,Fasano1999,Fasano1999b,Fasano2000,Grigorenko2001,Surdeanu2001,Silevitch2001,Field2002,Fasano2002,Menghini2003,vanBael2003,Fasano2003,Veauvy2004,Karapetrov2005,Bjornsson2005,Yurchenko2006,Fischer2007,Fasano2008,Petrovic2009,Suderow2014,AragonSanchez2019,Rumi2019,Llorens2020,Fisher1991,Nelson1993,Giamarchi1997,Civale1997,Fedirko2018}
Correlation function and structure factor studies have been
performed in order to characterize the structure of the different
glassy phases stabilized by disorder with different geometrical
properties, including randomly-distributed point
pins~\cite{Petrovic2009,AragonSanchez2019,Rumi2019} and correlated
defects,~\cite{Leghisa1993,Dai1994,CejasBolecek2016} as well as
periodic distributions of pinning sites.~\cite{Fasano2005} Recently,
some studies focusing on characterizing
long~\cite{Rumi2019,Llorens2020b} or short-range~\cite{Llorens2020}
vortex density fluctuations obtained contrasting results for samples
with point or correlated disorder, see the discussion in
Supplementary Note 1.  From the perspective of technological
applications, having information on the nature and distribution of
disorder in a given sample is crucial since correlated in addition
to point-like disorder is more efficient to pin vortices and thus to enhance the
material critical current.~\cite{Blatter1994}

\begin{figure*}[ttt]
\centering
\includegraphics[width=2\columnwidth]{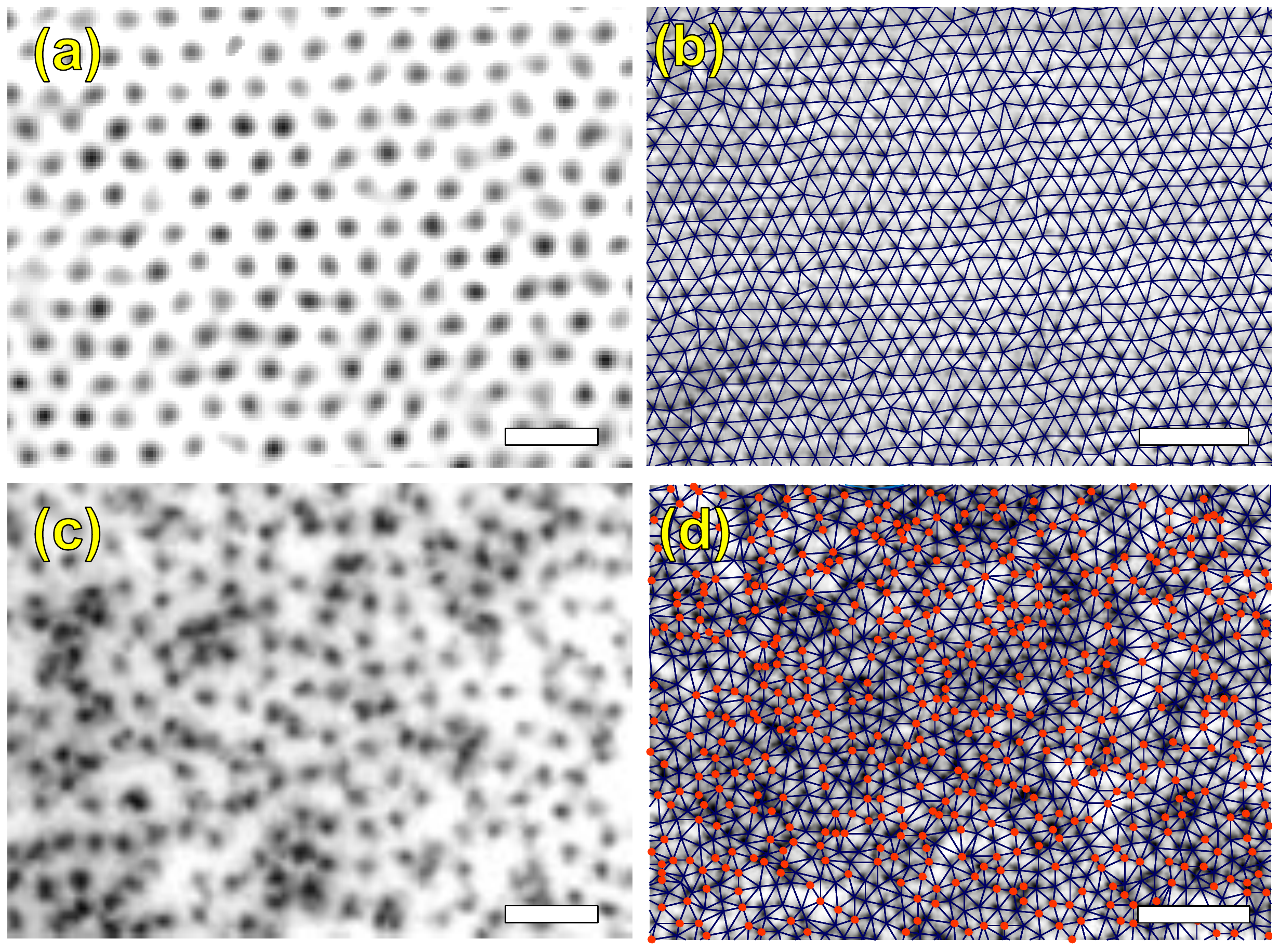}
\caption{Magnetic decoration snapshots of vortices (black dots)
nucleated at 30\,G in Bi$_2$Sr$_2$CaCu$_2$O$_{8+\delta}$
samples: (a), (b) Pristine sample with weak point disorder;  (c),
(d) sample with correlated disorder generated by a diluted density
of columnar defects with matching field $B_{\Phi}=30$\,G. Left
images cover the same field-of-view with the white bar corresponding
to 2\,$\mu$m; right images also cover the same but larger field of
view with the white bar corresponding to 5\,$\mu$m. The right panels
show superimposed Delaunay triangulations, an algorithm that
identifies first-neighbors, here joined with blue lines. Highlighted
in red are non-sixfold coordinated vortices, i.e. topological
defects in the hexagonal structure. These images are zoom-ins of
larger snapshots with approximately 7000 vortices.}
\label{fig:Figure1}
\end{figure*}

With the aim of inferring the nature of disorder in the host medium
from physical properties of the elastic objects, in this work we
follow a novel approach and study the inhomogeneous spatial
distribution of the particle-particle interaction force in extended
fields-of-view. Previous works by some of us combined the study of
interaction energy and force maps to infer information on the
typical separation between strong point disorder in pnictide
superconductors.~\cite{Demirdis2011,vanderBeek2012,Demirdis2013} But
here we go beyond that and study the more general problem of
discerning whether the host medium presents point or correlated
disorder. In this work, our case-study elastic system is the vortex
matter nucleated in the layered high-$T_{\rm c}$
superconductor Bi$_2$Sr$_2$CaCu$_2$O$_{8+\delta}$. We study samples
with different types of pinning landscapes representative of
different classes of  randomly distributed disorder:
Naturally occurring weak and dense \textit{point} pins in pristine samples,
extra moderate and dense \textit{point} pins generated by
electron irradiation, and columnar-defects (CD) responsible for \textit{correlated}
pinning. In the latter case, the pinning centers are columns of
crystallographic defects generated via heavy-ion irradiation, that
traverse the whole sample thickness and are distributed at random in
the plane perpendicular to the direction of vortices. We study samples with dilute and
dense distributions of correlated disorder quantified by the
matching field $B_{\Phi}=N_{CD} \cdot \Phi_{0}$ proportional to the
number of CD per unit area, $N_{CD}$, and the magnetic flux quantum
$\Phi_{0}=2.07 \cdot 10^{-7}$\,G$\,\cdot\,$cm$^2$. Our samples
with diluted CD disorder have $B_{\Phi}= 30, 45$ and $100$\,G,
whereas a sample with a larger $B_{\Phi}= 5000$\,G  is also studied.

 In the ideal case of mechanical
equilibrium, for a static configuration of vortices at a fixed
temperature, the vortex-vortex (i.e. the particle-particle)  interaction force  is
compensated by the vortex-pinning interaction. Therefore, the
spatial distribution of particle-particle interaction force allows for
the estimation of local pinning forces at the temperature at which
the snapshot was
captured.~\cite{Demirdis2011,Yang2012,vanderBeek2012,Demirdis2013,Yaguil2016,
CejasBolecek2016,AragonSanchez2019b} In our experiments we obtain
these snapshots  by decorating vortex positions with magnetic
nanoparticles  after following a field-cooling procedure (see
Methods for technical details). Even though the snapshots are taken
at 4.2\,K, the vortex structure observed by means of magnetic decoration
is frozen at temperature $T_{\rm freez}>>4.2$\,K given by disorder
inhibiting vortex motion at lengthscales larger than the vortex
spacing.~\cite{Pardo1997,CejasBolecek2016}
Once $T_{\rm freez}$
for a given host medium is determined (see the Supplementary Note 2
for details on how we measure  the field-evolution of $T_{\rm
freez}$), the particle-particle interaction force per unit length
for a given vortex $i$ with the rest of the $j$-th vortices of the
structure can be computed as~\cite{Blatter1994}

\begin{equation}
    \mathbf{f}_{\rm i}(\mathbf{r_{i}})=\frac{2\epsilon_0}{\lambda(T_{\rm freez})} \sum_{j}
     \frac{\mathbf{r}_{\rm ij}}{r_{\rm ij}}  K_{1}\left( \frac{r_{\rm ij}}{\lambda(T_{\rm freez})}\right)
     \label{fuerzaeq}
\end{equation}

\noindent This expression is valid for superconductors with large
$\kappa=\lambda/\xi$ values as Bi$_2$Sr$_2$CaCu$_2$O$_{8+\delta}$,
with $\lambda$ the in-plane penetration depth and $\xi$ the
coherence length, and in the low vortex density range $a_{0} \gg
\lambda$ covered in our experiments.  The magnitudes in
Eq.\,\ref{fuerzaeq} stand for: $\mathbf{r}_{\rm i}$ the location of
vortex $i$; $\mathbf{r}_{\rm ij}$ the vector separation between
vortices $i$ and $j$; $\epsilon_{0}= (\Phi_{0}/4\pi\lambda_{\rm
ab}(T_{\rm freez}))^2$ an energy scale proportional to the vortex
line energy  and $K_{1}$ the first-order modified Bessel
function.~\cite{Blatter1994} The sum runs for all vortices in the
sample but,  for the low-density vortex structures studied here, the
contribution from vortices separated more than $\sim 10 a_{0}$ is
negligible.

Maps of $\mathbf{f}_{\rm i}(\mathbf{r_{i}})$ depicting its spatial
variation  can be obtained from digitalizing vortex positions in
magnetic decoration snapshots as those shown in
Fig.\,\ref{fig:Figure1}. These images correspond to  zooms on structures of
$\sim 7000$ vortices  nucleated at 60\,G in media with point (top panels) and correlated
CD disorder (bottom panels). In the former case the vortex structure has
long-range orientational hexagonal order and there are no apparent short-scale
density fluctuations. In contrast, the vortex structure nucleated in the
medium with correlated
disorder presents noticeable degradation of the hexagonal symmetry
and strong short-scale vortex density fluctuations, with a tendency
to clustering at  some particular locations. Nevertheless, even if
the density of vortices in Figs.\,\ref{fig:Figure1}\,(c) and (d) is
equal to the global density of randomly distributed CD
($B_{\Phi}=30$\,G), not every vortex is located on top of a
correlated defect. This last assertion is proved in detail in the
Supplementary Note 3. Thus, the vortex clustering observed in
samples with dilute correlated disorder is connected to the spatial
distribution of CD's, but the structure imaged in the entire
field-of-view does not  mimic the random Poissonian
distribution of these correlated pins. Therefore, characterizing the
spatial distribution of vortices at densities $B/B_{\Phi}=1$ is not
an unambiguous way to ascertain whether the dominant disorder in the
medium  is point or correlated.

This contrast between the short-scale density-variation of vortex
structures nucleated in point- and correlated-disordered media  is
systematically found in larger field-of-view images, irrespective of
the vortex density. Indeed, at a given vortex density, the spatially
inhomogeneous distribution of first-neighbor distances $a$, is
larger for dilute correlated than for point disorder. The
Supplementary Note 4 presents the magnetic field
dependence of the standard deviation $SD$ of $a$ normalized by the
mean lattice spacing $a_{0}$  for all  studied samples. As
discussed there, despite this quantitative difference in $SD/a_{0}$,
the value and field-evolution of this magnitude is not a qualitative
indicator of disorder being point or correlated in nature. At best,
values of $SD/a_{0}>0.2$ might be taken as a hint that in this
particular vortex system the medium presents  dilute correlated
random disorder.

The panels (b) and (d) of Fig.\,\ref{fig:Figure1} also present
superimposed Delaunay triangulations that join first-neighbors with
blue lines.~\cite{Fasano2005} Topological defects formed by
non-sixfold coordinated vortices are highlighted in red. These
images are representative of results found in larger fields-of-view.
An analysis on the density of non-sixfold coordinated vortices,
$\rho_{\rm def}$,  in each type of medium reveals that for this
vortex density the structure nucleated in point disorder is
single-crystalline whereas the one nucleated in correlated disorder
is amorphous. However, at smaller vortex densities the $\rho_{\rm
def}$ enhances and even reaches similar values in structures
nucleated in point- and correlated-disordered media, see
Supplementary Note 4 for details. Then the value of $\rho_{\rm def}$
is neither a good candidate for distinguishing between point and
correlated disorder.

\begin{figure}[ttt] \centering
    \includegraphics[width=\columnwidth]{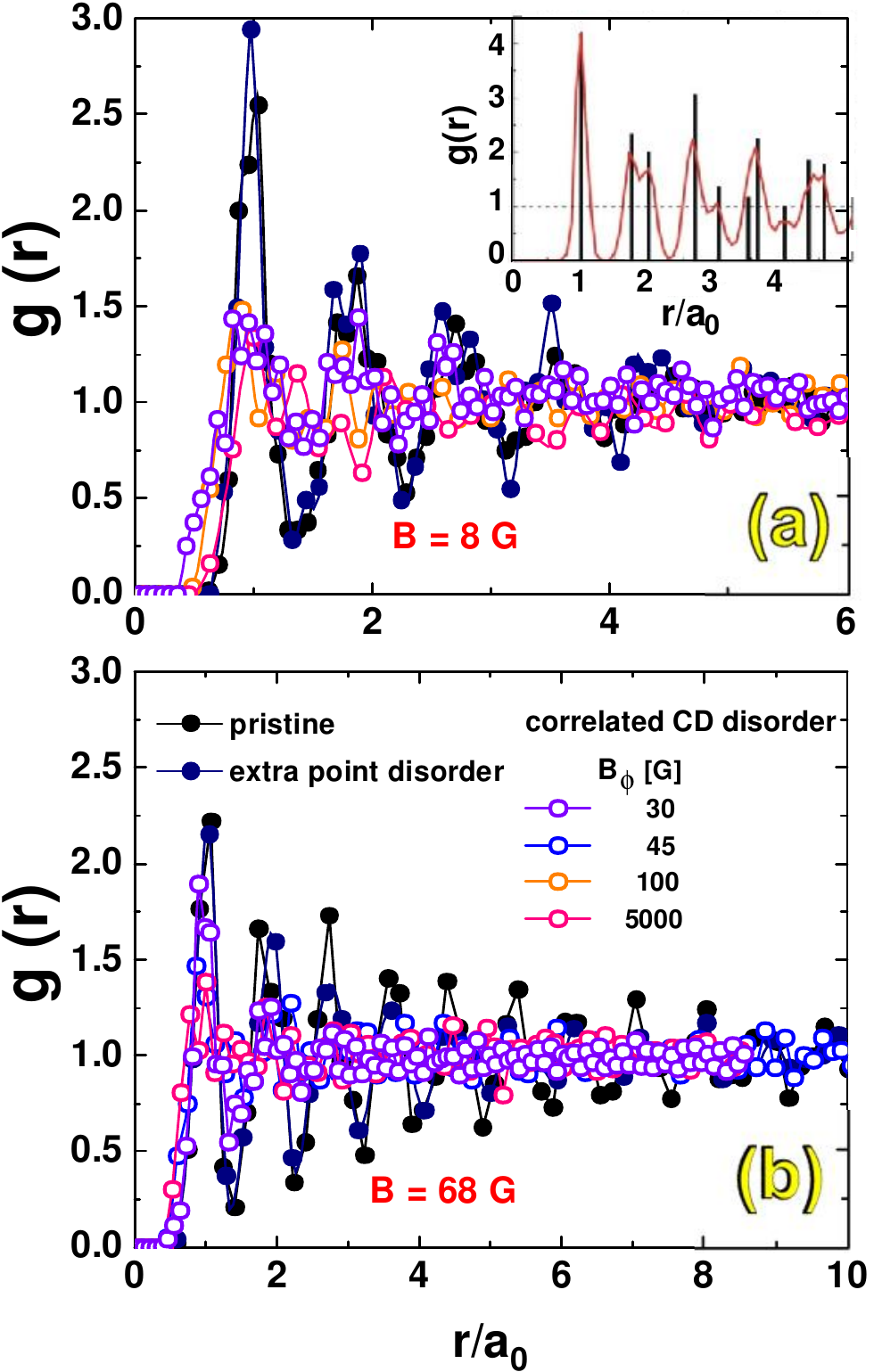}
    \caption{Pair correlation functions of vortex structures nucleated
        in Bi$_2$Sr$_2$CaCu$_2$O$_{8+\delta}$ samples with point (pristine
        and electron irradiated) and CD correlated (heavy-ion irradiated)
        disorder. Results at two different vortex densities of (a)
        8 and (b) 68\,G. The legend shows the color-code used
        for the different medium studied in both panels. Inset at the top: pair
        correlation function for a perfect
        hexagonal structure (black delta functions) and widening of the
        peaks due to random fluctuations around the sites of a perfect
        lattice (red curve).} \label{fig:Figure2}
\end{figure}

\begin{figure*}[ttt]
\centering
\includegraphics[width=2\columnwidth]{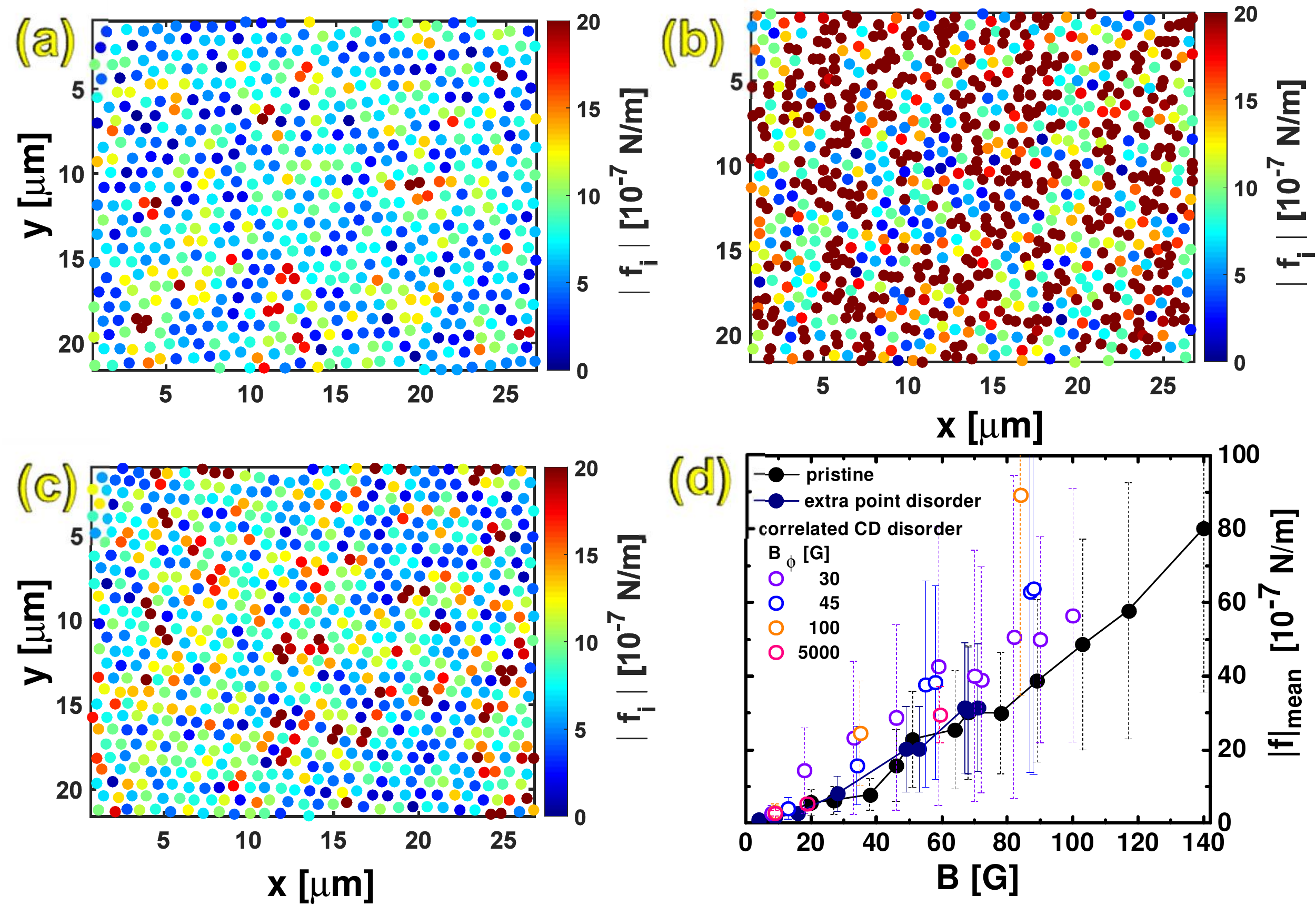}
\caption{Colour-coded maps of the modulus of the particle-particle
interaction force $|f_{\rm i}|$ for 30\,G vortex structures
nucleated in Bi$_2$Sr$_2$CaCu$_2$O$_{8+\delta}$ samples with point
((a) pristine, (c) electron- irradiated) and  correlated CD disorder
((b) heavy-ion irradiated with $B_{\Phi}=30$\,G). (d) Density
(field)-dependence of the mean value of $|f_{\rm i}|$, $|f|_{\rm
mean}$, obtained from large field-of-view maps with 7000-15000
vortices for all the studied structures. Bars accompanying the
points are not error bars but the value of the standard deviation of
the  $|f_{\rm i}|$ distribution registered in such extended
fields-of-view. \label{fig:Figure3}}
\end{figure*}

 Another possible path to reach this
ascertainment might be to study the pair correlation function $g(r)$
at different particle-to-pinning sites density-ratio. This function
describes how the density of particles varies as a function of
distance from a reference point. For a structure with mean lattice
spacing $a_0$, the $g(r)$ quantifies the probability of finding a
particle at a given distance $r/a_0$ from the origin, averaging this
probability when considering every particle of the structure at the
origin. This is therefore an angular-averaged probability that gives
information on short- and intermediate-distance vortex density
variations since $g(r) \to 1$ for $r$ larger than some units of
$a_{0}$. In the extreme cases of an ideal gas, $g(r)$ has a value of
1 independently of $r/a_0$, and for a perfect lattice presents delta
functions at distances corresponding to first neighbors, second
neighbors, and so on. The particular geometry of the lattice gives
the $r/a_0$ values at which these peaks are observed, see for
instance the inset of Fig.\,\ref{fig:Figure2}\,(a) for the case of a
perfect hexagonal structure. Disorder in the host sample shortens
the positional order of the structure and then induce a widening of
these delta functions.

Figure\,\ref{fig:Figure2} shows the $g(r)$ for vortex structures
nucleated at two different vortex densities of 8 and 68\,G, in
 three classes of disordered medium. In the case of point
disorder,  at the lower vortex density of 8\,G, $g(r)$ is quite
similar for structures nucleated in pristine and electron-irradiated
samples. However, on increasing field, sharper peaks are developed
in the case of  pristine samples, see Fig.\,\ref{fig:Figure2} (b).
On enhancing field there is a systematic increase of the number of
peaks observed in $g(r)$ for structures nucleated in both types of
point-disordered media, in accordance with vortex-vortex interaction
becoming more relevant with $B$. For a given vortex density, the
$g(r)$ for structures nucleated in samples with point disorder
presents several sharp peaks in contrast to a lesser number of peaks
detected for structures nucleated in a medium with correlated
disorder. This washing out of the peaks produced by correlated
disorder suggests in this medium pinning dominates over
vortex-vortex interaction and produces a substantial enhancement of
the displacement of particles with respect to the sites of a perfect
hexagonal lattice even at distances as short as $r/a_0\sim 2$.
However, using the number of peaks detected in $g(r)$ and its
sharpness seems not a categorical criterion to determine whether
disorder in the medium is dominated by point or correlated pins. For instance, for
the 68\,G structure nucleated in a sample with correlated disorder ($B_{\Phi}=30$\,G),
three peaks in $g(r)$ are clearly distinguished, a
phenomenology also observed  in samples with point pins at low
vortex densities.

We pursue our search of an unambiguous indicator in the physical
properties of the elastic structure for distinguishing correlated
from point disorder in the medium, by finding a magnitude that
should have a qualitatively different behavior for both types of
disorder. This property might be a particular feature of the
statistic distribution of the locally-varying particle-particle
interaction force that entails information on the short-scale density
fluctuations induced by disorder in the media.  With this aim we map
the vortex-vortex interaction force $\mathbf{f}_{\rm
i}(\mathbf{r_{i}})$ in extended fields-of-view following the
expression of Eq.\,\ref{fuerzaeq}. Maps of the modulus of the local
force, $|f_{\rm i}|$, are shown in Fig.\,\ref{fig:Figure3} for the
structures presented in Figs.\,\ref{fig:Figure1}, and also for
vortex lattices nucleated in samples with extra point disorder. In
all cases the vortex density corresponds to $B=30$\,G. There is
no noticeable spatial pattern in the $|f_{\rm i}|$ maps of vortex
structures nucleated in samples with point disorder, see panes (a)
and (c). On the other hand, clusters of bordeaux vortices with
larger modulus of the interaction force are distinguished in the
case of the correlated disordered medium, see panel (b). These
regions correspond to areas in which the vortices are closer than in
the rest of the structure, presumably induced by a locally denser
concentration of CD distributed at random in the sample. The mean of
the local values
 of $|f_{\rm i}|$, $|f|_{\rm mean}$, is plotted in
 Fig.\,\ref{fig:Figure3} (d) as a  function of field and for all the
 studied media.  The values of $|f|_{\rm mean}$, a  good estimate of the average
  pinning force,~\cite{Demirdis2011,Demirdis2013}   are
  always larger in vortex structures  nucleated in samples with strong correlated
  than with weak point disorder: Globally between
 30 to 50\,\% larger, and at high fields even reach a value 300\,\% larger for
 the particular  sample with $B_{\Phi}=100$\,G.
 The case of a medium with a dense  distribution of CD ( $B_{\Phi}=5000$\,G sample)
 is rather special since the $|f|_{\rm mean}$
  values  are close to those of samples with point disorder at low fields. However,
  at large $B$ the $|f|_{\rm mean}$ curve for the $B_{\Phi}=5000$\,G
  sample enhances significantly and tends towards the values found in the case
  of dilute correlated disorder. Then, due to their relative quantitative
  nature,  this global magnitude is not appropriate to ascertain the degree of
 correlation of the pinning sites.

\begin{figure*}[ttt]
\centering
\includegraphics[width=1.9\columnwidth]{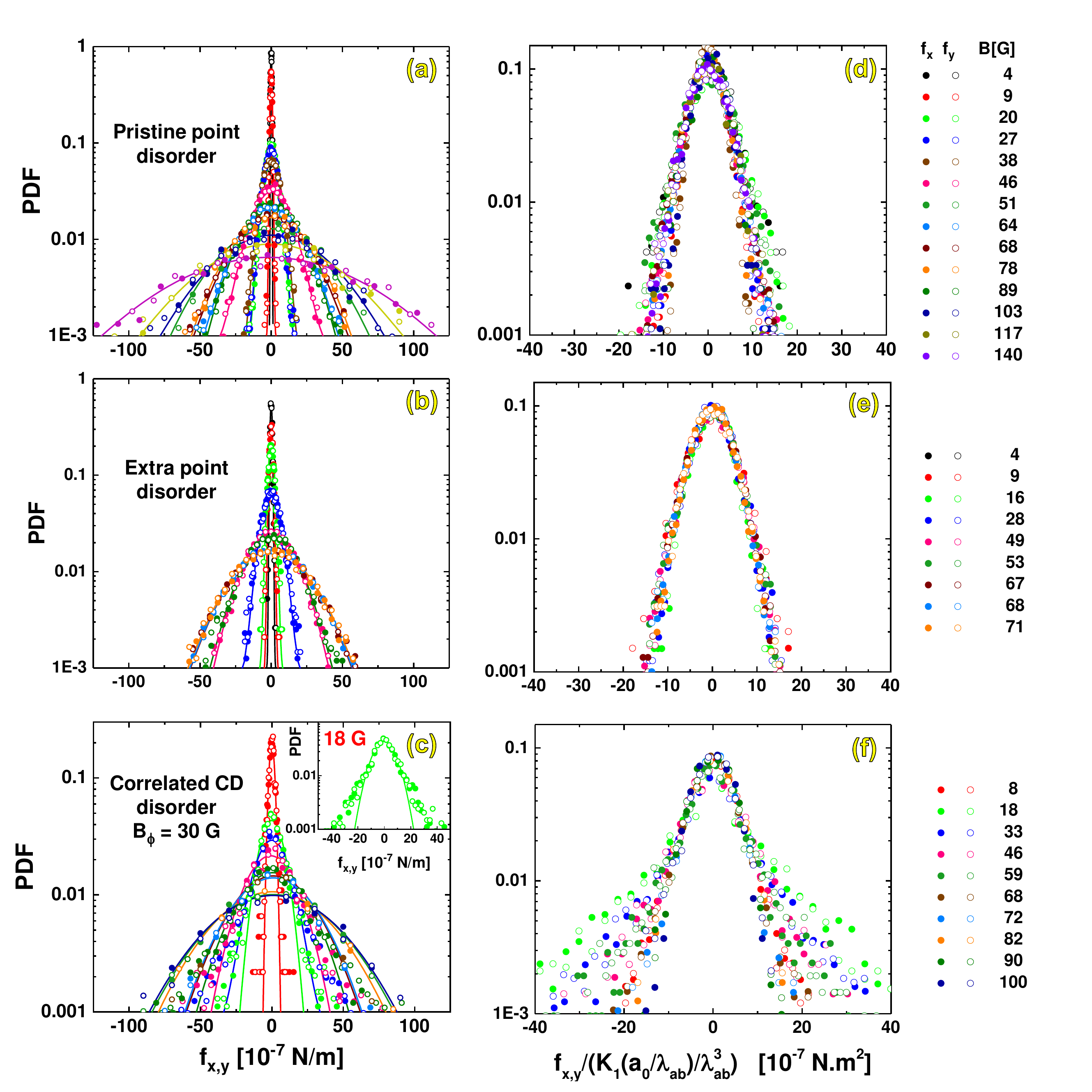}
\caption{Probability density functions (PDF's) of the
particle-particle interaction force components $f_x$ (full points)
and $f_y$ (open points) for vortex structures nucleated at various
applied fields in Bi$_2$Sr$_2$CaCu$_2$O$_{8+\delta}$ samples with
point (a) pristine, (b) electron irradiated, and (c) correlated CD
disorder with $B_{\Phi}=30$\,G. Data are presented in log-linear
scale. Full lines are fits to the data with a Gaussian
 function $\propto (1/\sigma_{\rm G}) \cdot
\exp(-x^2/2\sigma_{\rm G}^2)$ in the whole force range. This
function describe reasonably well the PDF's in the case of point
disorder, but in the case of correlated disorder fail to follow the
experimentally-observed tails that decay slowly with $f_{x,y}$ than
a Gaussian function. The inset in (c) shows the example of the
18\,G data where it is evident that the Gaussian function (full
green line) underestimates the experimental data in the large-force
range. The right panels (d) to (f) show the PDF's of the left panels
plot in log-linear scale with the $x$-axis normalized by the factor
$K_{1}(a_{0}/\lambda_{\rm ab})/\lambda^{3}$ and then normalized to
have an area under the curve equal to one. In this representation,
for point disorder the PDF's scale in a single evolution whereas for
dilute correlated disorder the curves do not overlap in the
large-force range. \color{black} \label{fig:Figure4}}
\end{figure*}

To grasp on this issue, we rely on magnitudes that significantly
depend on the local variations of vortex density that are presumably
larger and more inhomogeneous in the case of correlated disorder.
These magnitudes are the in-plane components of the interaction
force, $f_{x}$ and $f_{y}$, that we also map in extended
fields-of-view.  Examples of the probability density functions
(PDF's) for both components  are shown in Fig.\,\ref{fig:Figure4}
for structures nucleated in samples with point and dilute
correlated CD disorder ($B_{\Phi}=30\,$G in this case). These
data are some examples of the more than 50 cases studied covering
vortex densities between 4 and 140\,G and samples that are pristine, that have
extra point and dilute or dense correlated CD
disorder. In all our experimental data, the mode values of the PDF's
of $|f_{\rm i}|$ are finite (since the observed structures are far from being
a perfect hexagonal one), but the mode values of the PDF's of the
force components are equal to zero since the positive and negative
$x$ and $y$ directions of space are equivalent. Higher values of the
components of the particle-particle force enhance their probability
of occurrence on increasing $B$ since vortices get closer to each
other, irrespective of the type of disorder of the host medium.

Nevertheless, the right panels of Fig.\,\ref{fig:Figure4} show a
scaling of the data that highlight that the shape of the PDF for
large $f_{x,y}$ values is different for point than for dilute
correlated disorder. This scaling is made by dividing the force
components by a factor $K_{1}(a_{0}/\lambda(T_{\rm
freez}))/\lambda(T_{\rm freez})^{3}$ proportional to the average
interaction force for each studied field in a given material. While
for point disorder all curves collapse in a single trend, that is
not the case for dilute correlated disorder in the large-force
range. In the latter case, the tails of the scaled distributions do
not overlap and become narrower on increasing field, see
Fig.\,\ref{fig:Figure4} (f). Going back to the non-scaled data shown
in the left panels, the PDF's for structures nucleated in a medium
with point disorder fit a Gaussian distribution function
$(1/\sigma_{\rm G})\cdot \exp(-x^2/2\sigma_{\rm G}^2)$, see full
lines in (a) and (b).  In contrast, Fig.\,\ref{fig:Figure4} (c)
shows that for dilute correlated CD disorder a fit to the PDF's
with a Gaussian function (for the whole force range) only follows
the experimental data in the low-force range. In the large-force
range, the experimental data decay at a slower pace and non-Gaussian
tails are observed. This is clearly observed in the example shown in
the inset to this figure for the 18\,G vortex density. These
non-Gaussian tails become wider at low vortex densities, i.e. the
Gaussian fit is gradually underestimating the experimental data on
decreasing $B$. This finding  is also observed in the scaled-PDF's
of Fig.\,\ref{fig:Figure4} (f).

\begin{figure}[ttt]
\centering
\includegraphics[width=\columnwidth]{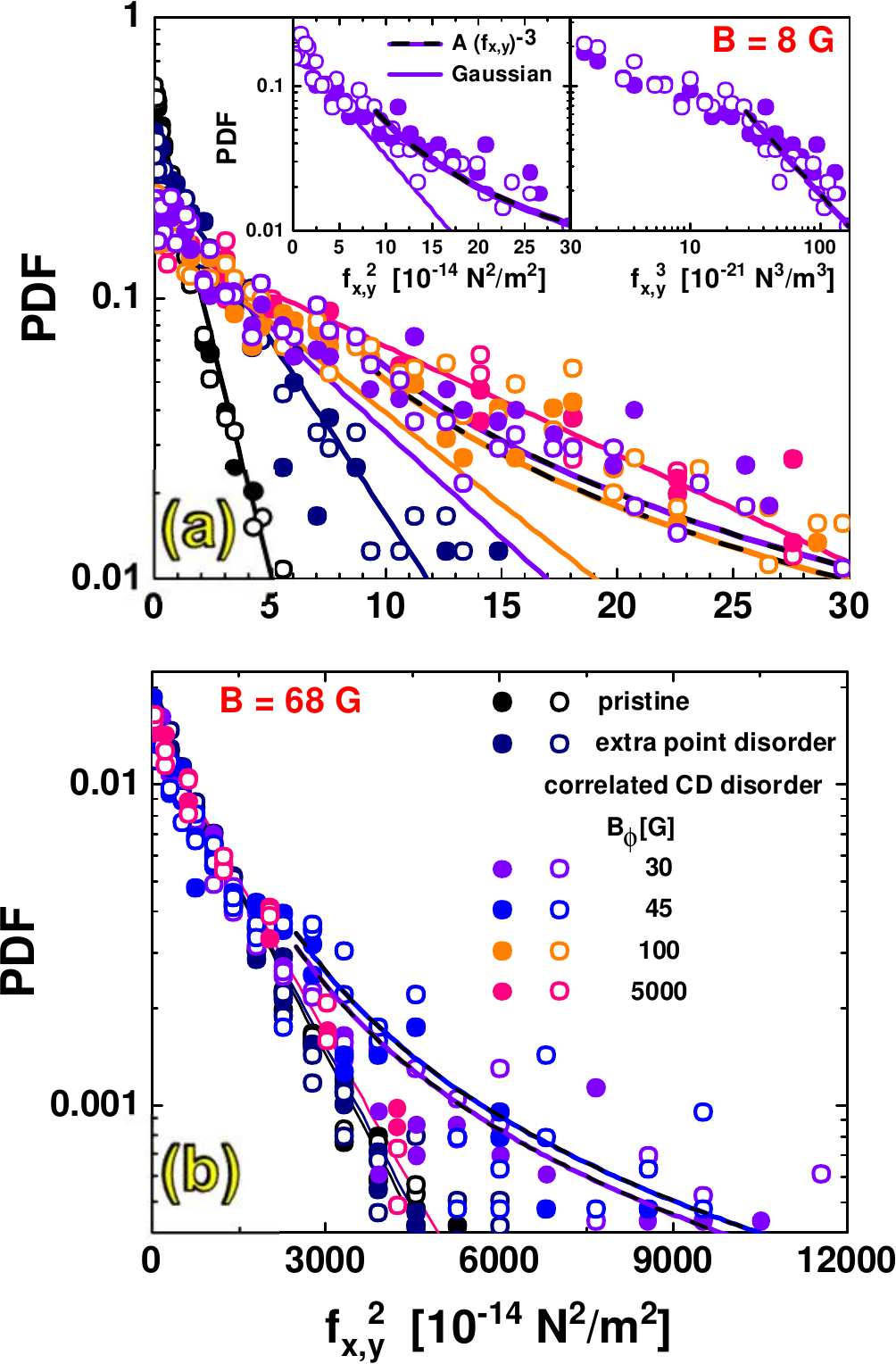}
\caption{Probability density functions (PDF's) of the vortex-vortex
interaction force components as a function of $f_{x}^2$ (full
points) and $f_{y}^2$ (open points) for densities of (a) 8 and (b)
68\,G for structures nucleated in samples with point and
correlated disorder with different CD densities. Fits to the data
with Gaussian functions (full lines) and with an algebraic decay
$\propto f_{x,y}^{-3}$ (black-dashed on top of color lines) are
presented. Insets: Detail of the 8\,G structure nucleated in a
sample with correlated disorder ($B_{\Phi}=30$\,G). Left: PDF
data plotted as a function of $f_{x,y}^2$ with Gaussian fit in full
line and algebraic fit in black-dashed on top of color line. Right:
Data plotted as a function of $f_{x,y}^3$ and the same algebraic
function shown in the left inset.} \label{fig:Figure5}
\end{figure}

Figure\,\ref{fig:Figure5} shows a different representation of some
examples of the studied cases including data in samples with point
and correlated CD disorder with different $B_{\Phi}$, both dilute
and dense. Panel (a) of this figure shows PDF data at a density of
8\,G and panel (b) shows data at 68\,G for all the studied
media. The data are plotted in a log-linear scale as a function of
$f_{x,y}^{2}$, a representation that puts in evidence that when the
host media present point disorder, the PDF's of the force
components follow a Gaussian distribution, irrespective of the
vortex density. The same figure shows that for a medium with diluted
CD disorder the PDF's of the force components follow a Gaussian
function only in the small-force range. However, in the
range of intermediate- and large-force components,
the PDF's plotted as a function of $f_{x,y}^{2}$ in log-linear scale do
not follow the apparent linear behavior. This is clearly illustrated
in the left inset to Fig.\,\ref{fig:Figure5} (a) showing the
example of the 8\,G data for a sample with $B_{\Phi}= 30$\,G: Data
departs from linearity (full
line) at intermediate values of the force components and its decay
is slower within the large-force range. In this  force-range where the
non-Gaussian tails develop, the PDF's are
well described by the algebraic decay $\propto f_{x,y}^{-3}$,
see for instance the apparent straight-line fit  shown on the
right insert in a log-linear representation as  a function of $f_{x,y}^{3}$.
This behavior is found for all the vortex structures nucleated in
samples with dilute correlated disorder irrespective of the vortex
density, but the force component value at which the departure
from the Gaussian behavior starts increases with $B$.  Data for
the structures nucleated in $B_{\Phi}=5000$\,G samples is again special:
The PDF's of $f_{x,y}$ follow an apparent linear
evolution with $f_{x,y}^2$ on log-linear representation, irrespective of the
vortex density up to 100\,G (see for instance the pink data
shown in Figs.\,\ref{fig:Figure5} (a) and (b)). Then, for a medium
with dense correlated disorder the distributions of the force
components are Gaussian as in the case of a host sample with weak and
dense point disorder. This finding suggests that,
in a more general perspective, Gaussian tails are expected for the PDF's of
the vortex-vortex force when pinning is in the weak limit (as is the case in
the $B_{\Phi}=5000$\,G samples~\cite{vanderBeek2000}) in contrast to the
algebraic tails detected in the case  of
dilute  correlated disorder  producing a strong pinning.
Data on superconducting samples with strong point pins could allow to
study this implication in the future.

 The non-Gaussian tails
observed in the PDF's of the vortex-vortex force components
originate in closely-located vortices, corresponding to the
cluster-like regions observed in samples with dilute CD correlated
disorder, see Fig.\,\ref{fig:Figure1} and the bordeaux vortices in the force
maps of Fig.\,\ref{fig:Figure3}. On the contrary, in the case
of point and dense correlated disordered media no tendency to vortex
clustering is observed and this might be at the origin of the PDF's
of the force components being well fitted by a Gaussian function in
the whole force range. Indeed, a closer look to the pair
distribution function of Fig.\,\ref{fig:Figure2} for $r/a_{0} < 1$
reveals that the probability of finding vortices at very small
distances $\sim 0.3-0.8\, r/a_{0}$ is larger in the case of
correlated disorder than of point disorder.

In an attempt to  explain how the spatial distribution of
the particle-particle interaction force entails the different short-scale
density variations in point and correlated disordered media, we
examine the force distribution that would be expected from model
spatial configurations. In general, the probability distribution
 of any of the components of the interaction force
between a given pair of vortices, $p(f^{pair}_{x})$, relates to the
probability density
$p'(r,\theta)$ of finding one vortex at the origin and another at a
position  $(r, \theta)$ (polar coordinates) by the expression
\begin{equation}
    p(f^{pair}_{x})=\int_0^\infty \int_0^{2\pi}p'(r, \theta)\delta(f^{pair}_{x}-\mathit{F}(r)\cos(\theta)) r d\theta
    dr,
    \label{Primera}
\end{equation}
\noindent where $\mathit{F}(r) \propto K_{1}(r/\lambda_{\rm
ab}(T_{\rm freez}))$ is the interaction force between any pair of
vortices separated a distance $r$. Note that this pair-force is not
the same as the component $f_x$ in Eq.\ref{fuerzaeq}, obtained as the sum
 of the interactions of one vortex with the rest. Nevertheless,
we expect that $p(f_{x}) \approx
p(f^{pair}_{x}=f_{x})$ in the limit of large $f_x$, as large
forces arise from few  very close neighbors. Therefore, both
distributions should have tails with the same shape.
While the number of vortex pairs
contributing to the tails of $p(f_x)$ is small,
the behavior of $p(f_x)$
at small $f_x$ is controlled by a large number
pair-forces proportional to $\sim (d/a_0)^2$, with
$d$ roughly the size of the field-of-view.
Since short-range
correlations are expected for most of these finite variance
contributions, we can invoke the
central limit theorem to assert that $p(f_x)$ presents a Gaussian
shape  around $f_x=0$. Then this explains why, regardless of the
type of disorder, the PDF's of $f_{x,y}$ follow a Gaussian behavior in the
low force range, see data on Figs.\,\ref{fig:Figure4} and \ref{fig:Figure5}.

\begin{figure}[ttt]
\centering
\includegraphics[width=0.9\columnwidth]{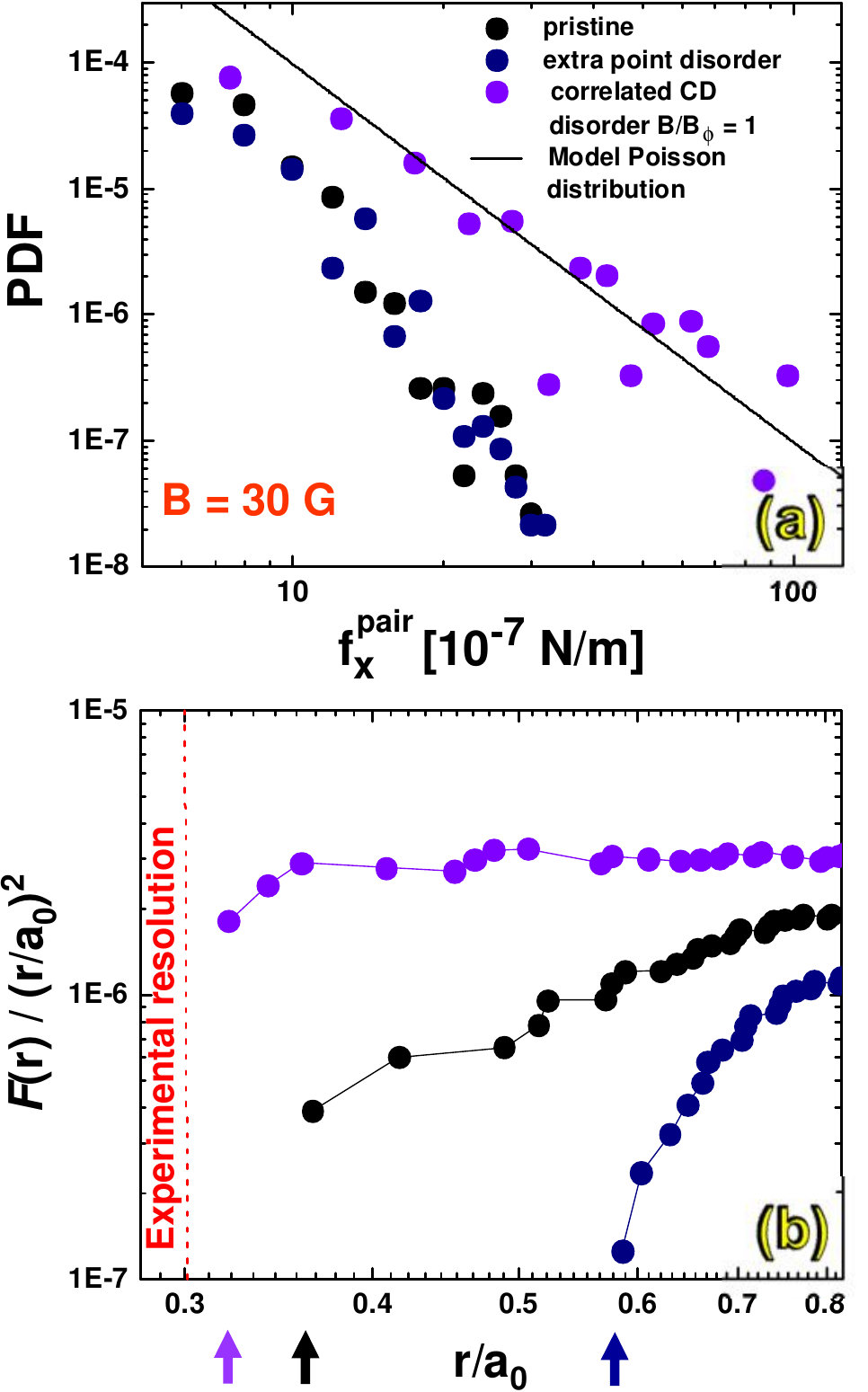}
\caption{(a) Probability density functions (PDF's) of the component
of the interaction force between pairs of vortices, $f^{pair}_{x}$,
for the vortex structures nucleated in samples with point (black and
navy points) and dilute correlated CD (violet points) disorder with
$B_{\Phi}=30$\,G. All the data correspond to a vortex density of
30\,G.  The analytical $1/f_{x}^3$ result for a toy-model
structure with a non-vanishing $g(r) \approx
\mathit{F}(r)/(r/a_{0})^2$ at small distances (such as a
vortex arrangement following a random Poissonian spatial
distribution) is shown with a black line. (a)
Normalized cumulative distribution function of the distance between
vortices in a pair, $\mathit{F}(r)/(r/a_{0})^{2}$, for the smallest
detected $r/a_{0}$ values for the same vortex structures studied in
panel (a). The threshold to resolve individual vortices with our
implementation of the magnetic decoration technique, $\sim
\lambda(4.2\,$K$)/a_{0}$, is indicated with a red dashed line.
Color-coded arrows located at the bottom indicate the
$r/a_{0}$ values corresponding to the smallest detected distance
between vortices in a pair in the whole field-of-view.
} \label{fig:Figure6}
\end{figure}

Let us now focus on the tails of the force distribution  and
consider the rather general fluid-like case of having an isotropic
vortex distribution. In this particular case, $p'(r, \theta) =  2\pi g(r)$ and
integrating Eq.\,\ref{Primera} over $\theta$ we get
\begin{equation}
    p(f^{pair}_{x})=\int_0^{\mathit{F}^{-1}(f^{pair}_{x})}
    \frac{4\pi r g(r)}{\mathit{F}(r)\sqrt{1-(f^{pair}_{x}/\mathit{F}(r))^2}}dr
    \label{Segunda}
\end{equation}
\noindent where $\mathit{F}^{-1}$ represents the inverse function of
$\mathit{F}$. Since  $\mathit{F}$ decreases monotonically with $r$,
it is therefore invertible, and then the integration limit is
uniquely defined. In order to analytically estimate the tails in the
$p(f^{pair}_{x})$ distributions, we consider the infinite
family of pair correlation functions that  rise as
$g(r) \sim r^{\alpha}$ for $r \ll a_0$, with  $\alpha \geq
0$ a characteristic exponent. In addition, we consider that
$\mathit{F}^{-1}(f) \sim 1/f$ for
large $f$ since $\mathit{F}(r) \propto K_1(r/\lambda) \sim 1/r$ for $r \sim \lambda \ll a_0$.
With these assumptions,
Eq.(\ref{Segunda}) can be integrated to obtain
\begin{equation}
    p(f^{pair}_{x}) \propto [f^{pair}_{x}]^{-(3+\alpha)}
    \frac{\Gamma \left(\frac{a}{2}+1\right)}{\Gamma \left(\frac{a+3}{2}\right)},
    \label{Tercera}
\end{equation}
\noindent where $\Gamma(x)$ is the Gamma function. We hence conclude
that for large forces $p(f_{x}=f) \sim p(f^{pair=f}_{x}) \propto
1/|f|^{(3+\alpha)}$ (using the equivalence of the tails of the PDF's at large $f$).
In particular, for a Poissonian ideal gas-like
distribution of particles we have $g(r)=1$ and then $\alpha=0$. The
prediction for this case is then $p(f_{x}=f) \sim p(f^{pair}_{x}=f)
\propto 1/f^{3}$. This result is however more general, as the same
result holds for any isotropic particle distribution with a non-zero
$g(r)$ at the smallest observable vortex-vortex distance. Most
often, for normal fluids with strong repulsive interactions, for $r \ll a_{0}$ $g(r)$
rises slower than any power law. In this case,
we can interpret that $\alpha=\infty$ effectively and then that
$p(f_{x}) \sim p(f^{pair}_{x})$ decays faster than a power-law for
large forces. Although these predictions are for large $f_x$ it is
important to keep in mind that they should be valid if
$f_x<\mathit{F}(r_{min})$, where $r_{min}$ is a cut-off distance.
This cut-off distance is given by the experimental resolution
to resolve individual vortices. In  magnetic
decorations, $r_{min} \sim \lambda(T)$, with $T$ the
temperature at which the experiment is performed.

We now check the  theoretical predictions described above by
comparing with our experimental data of the PDF's of $f^{pair}_{x}$ for
structures nucleated at 30\,G in samples with point (pristine
and electron irradiated) and dilute correlated CD ($B_{\Phi}=30$\,G) disorders,
see Fig.\,\ref{fig:Figure6} (a).  The
black curve corresponds to the analytical $1/f^3$ result found for a
Poissonian toy-model  structure with a
non-vanishing $g(r) \approx F(r)/r^2=cte$ for $r \ll a_{0}$. The figure
reveals that the vortex structure nucleated in a medium with dilute
correlated disorder displays a fair $1/(f^{pair}_{x})^3$ power-law
decay. In contrast, structures nucleated in pristine and extra point
disordered media display both a faster than power-law decay at large
$f^{pair}_{x}$. These findings are in  agreement with the tails of
the PDF's of the vortex-vortex interaction force components shown in
Fig.\,\ref{fig:Figure4}. This confirms the assumption that the PDF's of
$f^{pair}_{x}$ and $f_{x,y}$ share the same rare events statistics.

To test further the connection between the PDF's of the forces and
the distribution of  distances between vortices forming a pair ($g(r)$),
 we compute the cumulative distribution
of such distance, $\mathit{F}(r)\equiv \int_0^r dr'\;2\pi r' g(r')$,
for the smallest $r/a_{0}$ values detected experimentally. In order
to avoid spurious data binning effects for this rare events
statistics, we exploit the fact that the exact $\mathit{F}(r)$ can
be directly obtained from the data. To do this we first sort all
vortex-vortex distances from the smallest to the largest and use
them as the horizontal coordinate. We thus obtain the exact
$\mathit{F}(r)$ for the discrete data-set by computing the order-number
divided by the total number of vortex pairs. If we now model the rising
$g(r)\sim r^{\alpha}$, we have $\mathit{F}(r)\sim r^{2+\alpha}$.
Therefore, $\mathit{F}(r)/(r/a_{0})^2 \sim (r/a_{0})^\alpha$ gives
us access to the effective exponent $\alpha$ which also controls the
decay of the PDF's of the interaction force. Figure \,\ref{fig:Figure6}(b)
 shows $\mathit{F}(r)/(r/a_{0})^2$ for
the 30\,G vortex structures nucleated in media with weak point
 and dilute strong correlated disorder.
In the latter case, the cumulative distribution of
distances between vortices in a pair displays an almost flat
behavior down to the minimum value of this magnitude detected
experimentally (indicated with a violet arrow). Moreover,  no
tendency to a steep decrease of $\mathit{F}(r)/(r/a_{0})^2$  is
observed on decreasing $r/a_{0}$. This means that
$\alpha=0$ effectively, in fair consistence with the $1/(f_{x,y})^3$
 force distribution tails detected in structures
nucleated in samples with dilute correlated disorder. On the other
hand, for structures nucleated in point disordered media,
$\mathit{F}(r)/(r/a_{0})^2$ displays a faster-than-algebraic decay
on decreasing $r/a_{0}$, with a faster decay for the sample with
extra point disorder than for the pristine one. In both cases
the minimum value of distance between vortices in a pair
observed in the entire field-of-view is well above the
experimental resolution, indicating that $\alpha=\infty$
effectively. The prediction for point disordered media is then a
faster than algebraic decay of the force distributions, in agreement
with the Gaussian-shaped PDF's observed experimentally.

In summary, we present an alternative way on inferring the nature of
the dominant disorder present in the media where elastic objects are
nucleated based on the analysis of physical properties  of the
interacting elastic objects that can be computed from direct imaging
of the structures in fields-of-view containing a statistically
meaningful number of particles. We illustrate our proposal using
experimental data on vortex lattices in superconducting materials as
a case-study system. We analyze the statistical distribution of the
disorder-induced spatially-varying particle-particle interaction
force and found a  behavior distinctive for strong dilute correlated as opposed to
weak point-like disorder.
We show that detecting non-Gaussian algebraically-decaying tails in the PDF
of the components of the interaction forces acting on individual vortices
is a smoking gun proof of the randomly distributed disorder, in
our case dominated by dilute correlated defects acting as strong pinning
centers. This result contrasts with the Gaussian PDF's of the force components
 for structures nucleated in  media with point or very dense correlated disorder.
  By considering a toy-model system we
explain that the non-Gaussian tails result from inhomogeneous short-scale
vortex density fluctuations associated to the tendency of clustering in
some patches of the structure.  Whether our method
 is effective to distinguish between the host media presenting strong or weak disorder
 in a more general perspective remains as an interesting open question for
 further investigations. Nevertheless, our proposal is a very
 promising way of inferring the
nature of disorder in the host media of elastic objects from
physical properties of the structures directly imaged. Its
applications can be easily spanned to a wide range of
soft condensed matter systems in which distinguishing the nature of
disorder might be crucial for technological applications.

\section*{Methods}

The studied samples are nearly optimally-doped single crystals of
Bi$_2$Sr$_2$CaCu$_2$O$_{8+\delta}$ from different sample growers,
with natural and introduced defects distributed at random. We
studied a set of roughly 40 samples  grown by means of the
traveling-solvent-floating-zone~\cite{Li1994} and flux methods and
having $T_{\rm c} \sim 90$\,K.~\cite{Correa2001}. While some of
these samples were kept pristine, others were exposed to different
doses and types of irradiation. One underdoped sample was irradiated with
electrons with an energy of $2.3$\,MeV and a dose of $1.7 \cdot
10^{19}$\,e/cm$^2$ at the \'Ecole Polytechnique, France. The induced
damage by this irradiation resulted in  extra point disorder,
reduced the critical temperature of the sample down to 66\,K, and
raised $\lambda$ by roughly 30$\%$. ~\cite{Konczykowski2009}
Correlated CD disorder was generated by irradiating other pristine
Bi$_2$Sr$_2$CaCu$_2$O$_{8+\delta}$ samples with heavy-ions at the
GANIL facility in France. Some samples were irradiated with 6\,GeV
Pb-ions at corresponding matching fields of $\mathrm{B}_{\Phi}=45$,
100 and $5000$\,G, and others with  5\,GeV Xe-ions with
$\mathrm{B}_{\Phi}=30$\,G.
 Heavy-ion irradiation produced a
random poissonian distribution of CD parallel to the c-axis of the
sample. In these samples there was a negligible depression of the
critical temperature and no significant change in the value of
$\lambda(0)$.~\cite{vanderBeek2000}

Snapshots of the vortex structure at the surface of the sample are
obtained by performing magnetic decoration experiments at 4.2\,K
after a field-cooling process.\cite{Fasano2008} During this process
the vortex structure gets frozen  at length-scales of the lattice
parameter $a_0$ at a temperature $T_{\mathrm{freez}}$ and on further
cooling down 4.2\,K  vortices move in lengthscales of the order
$\xi$, much smaller than the typical  size of a vortex detected by
magnetic decoration, of the order of $\lambda$. Therefore the
structure imaged in such magnetic decoration experiments corresponds
to the equilibrium one at $T_{\rm{freez}}$. At this crossover
temperature the bulk pinning dominates over the vortex-vortex
repulsion and the thermal fluctuations.~\cite{Pardo1997,Fasano1999}
Then $T_{\rm{freez}}$ depends not only on the  superconducting
material but also on the particular pinning landscape and the
magnetic induction $B$. We estimate $T_{\rm freez}$ as of the order
of the irreversibility temperature $T_{\mathrm{irr}}$ at which bulk
pinning sets in on cooling.

In order to obtain $T_{\rm irr}(B)\sim T_{\rm freez}$ for each
particular sample, we measure the irreversibility line  by means of
local Hall probe magnetometry using micrometric Hall sensors with
active areas of $16 \times 16$\,$\mu$m$^{2}$.~\cite{Dolz2014} The
irreversibility temperature is taken at the onset of the non-linear
magnetic response due to the growing relevance of bulk pinning on
cooling. This onset is detected by measuring the vortex magnetic
response to an ac ripple field superimposed to the external static
magnetic field $H$, both parallel to the c-axis of the
sample.~\cite{Dolz2014} By applying a lock-in technique, the
response of the sample at the third harmonic of the excitation field
is recorded as a function of temperature and normalized as to obtain
the transmittivity $\mid T_{\rm h3} \mid$. This magnitude is zero in
the normal state and starts to have a finite value on cooling at the
temperature at which pinning sets in, namely $T_{\rm
irr}(B)$.~\cite{Dolz2014,Dolz2015,CejasBolecek2016}  Measurements
were typically performed with a ripple field of 1\,Oe and 7.1\,Hz.

\section*{Data availability}

All relevant data are available from the authors upon request.

\section*{Acknowledgements}

This work was supported by the ECOS-Sud-MINCyt France-Argentina
bilateral program under Grant A14E02; by the Argentinean National
Science Foundation (ANPCyT) under Grants PRH-PICT 2008-294, PICT
2011-1537, PICT 2014-1265 and PICT 2017-2182; by the Universidad
Nacional de Cuyo research grants 06/C566-2019 and
80020160200046UN-2016; and by Graduate Research fellowships from
IB-CNEA for J. P. and from CONICET for J. A. S., R. C. M., G. R. and
N. R. C. B. We thank to M. Li for growing some of the studied
pristine single crystals and V. Moser for providing us the Hall
sensors.

\section*{Author contributions}
Y. F. designed research; C.J. v.d.B, Y. F. , and A. K discussed the
general method to analyze the data, J. A. S., R. C. M., Y. F., N. R.
C. B., G. R., J. P., P. P., M. I. D. and M. K. performed
measurements, G. N. grew  pristine samples, M. K. and C. J. v. d. B
irradiated samples; J. A. S., R. C. M., N. R. C. B., G. R., J. P.,
P. P., A. B. K. and Y. F. analyzed data; G. R. and A. B. K.
performed theoretical calculations;  Y. F., A. B. K., J. A. S. and
G. R. wrote the paper.

\section*{Competing interests}
The authors declare no conflict of interest.

\section*{Materials and correspondence}
\textsuperscript{*}To whom correspondence should be addressed.
E-mail: yanina.fasano at cab.cnea.gov.ar

\end{document}